\documentclass[sigconf]{acmart}
\AtBeginDocument{%
  }

\setcopyright{rightsretained}
\copyrightyear{2025}
\acmYear{2025}
\acmDOI{XXXXXXX.XXXXXXX}
\acmConference[Conference acronym 'XX]{Make sure to enter the correct
  conference title from your rights confirmation email}{June 03--05,
  2018}{Woodstock, NY}
\acmISBN{978-1-4503-XXXX-X/2018/06}

\usepackage{multirow}
\usepackage{pifont}

\usepackage{color, colortbl}
\definecolor{LightCyan}{rgb}{0.88,.94,1}

\acmSubmissionID{7917}

\begin{document}

\title{Governing Together: Toward Infrastructure for Community-Run Social Media}

\author{Sohyeon Hwang} 
\orcid{0000-0001-8415-7395}
\affiliation{
  \institution{Princeton University}
  \city{Princeton}
  \state{NJ}
  \country{USA}} 

\author{Sophie Rollins} 
\authornote{Authors contributed equally and order in author list was randomized.}
\orcid{}
\affiliation{
  \institution{Northwestern University}
  \city{Evanston}
  \state{IL}
  \country{USA}} 

\author{Thatiany Andrade Nunes} 
\authornotemark[1]
\orcid{}
\affiliation{
  \institution{Northwestern University}
  \city{Evanston}
  \state{IL}
  \country{USA}} 

\author{Yuhan Liu} 
\orcid{}
\affiliation{
  \institution{Princeton University}
  \city{Princeton}
  \state{NJ}
  \country{USA}} 

\author{Richmond Wong} 
\orcid{}
\affiliation{
  \institution{Georgia Tech}
  \city{Atlanta}
  \state{GA}
  \country{USA}} 

\author{Aaron Shaw} 
\orcid{}
\affiliation{
  \institution{Northwestern University}
  \city{Evanston}
  \state{IL}
  \country{USA}} 

\author{Andrés Monroy-Hernández} 
\orcid{}
\affiliation{
  \institution{Princeton University}
  \city{Princeton}
  \state{NJ}
  \country{USA}} 

\renewcommand{\shortauthors}{Authors et al.}

\begin{abstract}
Decentralizing the governance of social computing systems to communities promises to empower them to make independent decisions, with nuance and in accordance with their values. Yet, communities do not govern in isolation. Many problems communities face are common, or move across their boundaries. We therefore propose designing for {\em inter-community governance}: mechanisms that support relationships and interactions between communities to coordinate on governance issues. Drawing from workshops with 24 individuals on decentralized, community-run social media, we present six challenges in designing for inter-community governance surfaced through ideas proposed in workshops. Together, these ideas come together as an ecosystem of resources, infrastructures, and tools that highlight three key principles for designing for inter-community governance: {\em modularity}, {\em forkability}, and {\em polycentricity}. We end with a discussion of how the ideas proposed in workshops might be implemented in future work aiming to support community governance in social computing systems broadly.
\end{abstract}



\maketitle

\section{Introduction} 
Community governance has been crucial to many social computing systems, ranging from discussion fora and local groups like Reddit and Facebook Groups, to peer production projects such as Wikipedia \citep{seering_reconsideringselfmoderation_2020, li_measuringmonetary_2022, benkler_commonsbasedpeer_2006}. 
Motivated by the potential of communities to govern with greater nuance, context, and care, efforts to improve community governance have typically focused on dynamics and interventions {\em within} communities. 

However, users, content, and interactions typically move across community boundaries. Neither the problems communities deal with nor the governance practices they devise in response are purely endogeneous. For example, a malicious actor creating multiple accounts to spam many communities at once is a broader problem that cannot be addressed by improving the governance of one community alone. Although there is growing empirical evidence of inter-community and ecological dynamics shaping community success \citep[e.g.,][]{teblunthuis_ecologyonline_2021,kiene_relationalorigins_2025,hortaribeiro_deplatformingnormviolating_2025}, far less work designs systems that would help communities manage and leverage these aspects of how they operate. In this work, we suggest that {\em inter-community} governance --- mechanisms that intervene in the relationships between communities to shape community governance --- is a crucial but undersupported area of design and systems-building that would strengthen community governance and make it more sustainable. Leaders of online communities already find {\em ad hoc} ways to learn from one another's communities \citep{kiene_relationalorigins_2025}, coordinate on widespread issues \citep{seering_prideprofessionalization_2022,matias_goingdark_2016}, and/or protect themselves from each other \citep{bono_explorationdecentralized_2024,jhaver_onlineharassment_2018,wu_negotiatingsociotechnical_2024}. At present, social computing systems are rarely designed to support this type of interaction {\em between} communities; admins and moderators thus put extra care and resources into leveraging existing affordances and devising workarounds.

To explore the design space of {inter-community} governance, we present findings from four design workshops with 24 community organizers and developers in the Fediverse, a decentralized, community-run social media network. On the Fediverse, independently operated and governed communities often have bearing on one another because members of different communities interact, even if those communities have divergent goals, norms, and so on. This results in communities frequently experiencing what \citet{hwang_trustfriction_2025} call  {\em governance frictions}: incompatibilities in the governance decisions of two communities that undermine the goals of one community in the course of advancing the goals of another. Drawing from workshop discussions about {\em ideal workflows} for addressing governance frictions, we identify six challenges for coordinating governance in community-run social media around information sharing, managing relationships, and encouraging adoption. The ideas proposed by participants envision an ecosystem of interfaces, infrastructures, and resources that emphasizes three key design principles: modularity, forkability, and polycentricity. We present the challenges, ideas, and principles as conceptual tools that future work can leverage as a road map for designing for inter-community governance in social computing. 

\section{Background} 
We begin by explaining the importance of community-level decision-making in social computing systems like social media. We then reflect on approaches for improving community governance in HCI literature, particularly in social computing, going from looking within communities to looking across, toward inter-community governance. We describe how prior work orients our attention to how inter-community governance might be supported. 

\subsection{Toward Community-Run Social Media} \label{sec:bg_value}
We define communities as discrete units of collective decision-making within social computing systems~\citep{foote_onlinecommunities_2023,bruckman_newperspective_2006}, where a self-defined group of users freely and voluntarily create and join a shared space toward common interests and have mechanisms to enforce shared values. Community governance thus refers to how these groups self-govern within a system. We focus on community governance in social media systems in particular, which we define per \citet{zhang_formfromdesign_2024}: ``systems that facilitate the sharing of user-generated expressive content, or media, from one user to others.'' This includes centralized platform systems such as Instagram, Pinterest, and Slack, as well as decentralized protocol-based systems such as Visual IRC, Bluesky, and Mastodon. 

Across social media systems, community governance has been a critical part of how governance online is enacted, providing voluntary but significant ``civic labor'' \citep{matias_civiclabor_2019}.  \citet{li_measuringmonetary_2022} estimate that the monetary value of volunteer moderation by communities on Reddit in 2020 was approximately 3.4 million USD (a conservative estimate). Qualitative work offers rich accounts of the emotional labor \citep{dosono_moderationpractices_2019} and care \citep{yu_takingcare_2020} community governance involves, across multiple dimensions of crafting a community space \citep{morgan_teasympathy_2013,narayan_wikipediaadventure_2017,seering_metaphorsmoderation_2020}. 
Because social media operates at ``unfathomable'' scales \citep{gillespie_scalejust_2018}, community-level governance can provide nuanced and contextualized decision-making that supports interactions across a much richer array of goals, topics, and values \citep{chandrasekharan_internetshidden_2018, weld_whatmakes_2021}. In this way, community governance can also be an opportunity for individuals to shape the norms of digital life \citep{gibson_freespeech_2019}, nurturing spaces they find valuable \citep{yu_takingcare_2020}. 

Amidst persistent concerns about the outsized power of centralized social media platforms~\citep{douek_contentmoderation_2022,odonnell_havewe_2021,denardis_internetgovernance_2015,hinds_itwouldnt_2020}, interest in building, understanding, and strengthening {\em community-owned and -run social media} systems has grown~\citep{sahneh_dawndecentralized_2024,cohn_fediversecould_2022,mansoux_seventheses_2020,masnick_protocolsnot_2019}. 
Community-run social media decentralizes power through shared, open protocols that enable an ecology of interoperable platforms \citep{masnick_protocolsnot_2019} rather than a centralized entity that provides oversight and constrains communities. For example, on the Fediverse --- a decentralized social media network that is the empirical focus of this work (introduced at length in \S \ref{sec:setting}) --- communities of users can create or join servers that enable them to interact with others through a shared protocol. Each server self-governs, offering users more control \citep{cohn_fediversecould_2022}. 
However, community governance comes with its own challenges \citep{hwang_trustfriction_2025,anaobi_willadmins_2023}. 
In addition to demanding care and labor \citep{dosono_moderationpractices_2019,li_allthats_2022}, community governance can require knowledge people do not have or even know to anticipate \citep{ tosch_privacypolicies_2024}. Community governance can thus be difficult to sustain, leading to psychological distress and burnout \citep{schopke-gonzalez_whyvolunteer_2024}. 
\subsection{Strategies and Tools for Community Governance} \label{sec:bg_success}

\subsubsection{Looking Within Communities}
\label{sec:bg_success_intra}
Quantitative metrics for improving community governance often measure activity or outcomes within a community, such as growth \citep{warncke-wang_increasingparticipation_2023}, turnover \citep{dabbish_freshfaces_2012}, compliance \citep{matias_preventingharassment_2019}, and toxicity \citep{jhaver_evaluatingeffectiveness_2021}. Accordingly, researchers and designers have focused on the internal dynamics that appear to lead to desirable outcomes \citep{kraut_buildingsuccessful_2012}. Prior work articulates the practices of moderators, admins and other leaders within a community \citep{lo_whenall_2018,seering_metaphorsmoderation_2020,squirrell_platformdialectics_2019,anaobi_willadmins_2023}, such as their efforts to foster safe spaces \citep{wu_negotiatingsociotechnical_2024,dosono_moderationpractices_2019, evans_facebookvenezuela_2018} or onboard newcomers \citep{ren_buildingmember_2012,morgan_teasympathy_2013}. Related work focuses on understanding the organizational and institutional dynamics of how communities govern in these socio-technical environments \citep[e.g.,][]{jiang_moderationchallenges_2019,schneider_adminsmods_2022,frey_emergenceintegrated_2019}. 
Researchers have additionally looked to devise and evaluate interventions (e.g., introduction of a new moderation approach) in a community \citep{tran_risksbenefits_2022,teblunthuis_effectsalgorithmic_2021}. Others involve building novel tools that aim to make community governance more sustainable by: reducing the volume of moderation work \citep{jhaver_designingword_2022,atreja_appealmodinducing_2024}, mediating conflict between members \citep{doan_designspace_2025}, automatically providing context for decisions \citep{kuo_unsungheroes_2023,song_modsandboxfacilitating_2023}; and scaffolding policy-making \citep{zhang_policykitbuilding_2020}.

The literature predominantly seeks to assess dynamics, mechanisms, and interventions {\em within} a community \citep{kraut_buildingsuccessful_2012}. Yet, many of the issues these interventions respond to are neither specific nor endogeneous to a particular community. They often reflect ubiquitous organizational problems that communities encounter, such as around technology adoption \citep{kiene_technologicalframes_2019} or improving rule compliance \citep{matias_preventingharassment_2019}.
Other times, they reflect broader problems on the social media, such as toxicity \citep{massanari_gamergatefappening_2017}, harassment campaigns \citep{han_hateraids_2023}, or other communities acting in bad faith \citep{kharazian_governancecapture_2023,heslep_mappingdiscords_2021}. In short, communities neither encounter nor address these issues in isolation; examining dynamics {\em within} a community alone is valuable but insufficient to address challenges communities face. 

\subsubsection{Looking Between Communities} \label{sec:bg_success_inter}
Prior work emphasizes ecological relationships between communities \citep{butler_membershipsize_2001,butler_crosspurposescrossposting_2011,lopez_consequencescontent_2013,wang_impactmembership_2012,teblunthuis_densitydependence_2017,teblunthuis_nocommunity_2022,teblunthuis_identifyingcompetition_2022,hwang_whypeople_2021},\footnote{Some work considers dynamics between communities across systems \citep[e.g.,][]{hortaribeiro_platformmigrations_2021,phadke_manyfaced_2020}; in this work, we focus on dynamics between communities within a given system.} beginning with the observation that communities overlap in topic and users. 
Although the fact that humans have limited attention might suggest competition among communities, \citet{teblunthuis_ecologyonline_2021} provides evidence of widespread mutualism instead, with communities often complementing one another \citep{teblunthuis_nocommunity_2022, hwang_whypeople_2021}. 

The community ecologies work tends to focus on participation. Related empirical work shows that communities also impact one another's governance practices. For example, evaluations of {\em deplatforming} by platforms from the top-down have found that moderation interventions in one community or system have direct, causal effects on the dynamics of others, albeit with some mixed substantive results \citep{hortaribeiro_deplatformingdid_2023,mekacher_systemicimpact_2023,ali_understandingeffect_2021,hortaribeiro_deplatformingdid_2023}. Even when communities don't directly engage with one another, they can experience {\em spillover effects} \citep{russo_strangerdanger_2024}. 
Communities may also {\em sanction} one another (e.g., community-level blocking), which can reduce activity in the sanctioned community \citep{colglazier_effectsgroup_2024}. Conversely, communities may want to {\em cooperate or collaborate} on issues, such as sharing strategies, setting boundaries, or coordinating against a bad actor. Research already documents how communities look to one another for governance issues. For example, recent work by \citet{kiene_relationalorigins_2025} describes how online communities devise their rules in {relational ways}: looking to other communities that they copy from, distinguish themselves from, and assert their boundaries against. 

These studies illustrate how the problems that a community deals with, and the governance practices they adopt, are inextricably tied to what happens in other digital spaces. We suggest that scaffolding {\em inter-community governance} --- strategies and tools that support relationships and interactions between communities for governance --- can create opportunities for communities to learn from one another, cooperate, and collaborate. 
Intuitively, such interactions may also help communities understand when and how to set boundaries against other communities that are incongruent with their goals or values, as well as negotiate and communicate with them. Opportunities for coordination are likely able to help communities set well-established governance processes, share labor in more sustainable ways, and exchange strategies toward improving outcomes and reducing costs. 

\subsection{How Might We Design for Inter-community Governance?} \label{sec:bg_design}
Despite the recognition that communities do not govern in isolation, little attention has gone to understanding and developing tools of inter-community governance. A notable exception is \citet{chandrasekharan_crossmodcrosscommunity_2019}'s {\em CrossMod}: an AI-backed system that provides recommendations for content moderation by leveraging a corpus of previous moderator decisions across communities, which the authors refer to as {\em cross-community learning}. However, {\em CrossMod} is primarily meant to shape decisions within a community. This tool's pooling of information from multiple communities thus only reflects a small piece of the kind of interactions that inter-community governance suggests. For example, \citet{seering_prideprofessionalization_2022} document how a campaign to coordinate volunteer moderators across Discord communities during Pride month successfully distributed additional moderation labor to LGBTQ+ communities who were targets of harassment raids. Although this case demonstrates the benefits of coordination across communities, the centralized nature of the campaign (organized by Discord) renders the design implications for doing so in a bottom-up manner unclear. 

Below, we briefly orient our approach to understanding how inter-community governance can be supported.

\subsubsection{Institutional Design}
We draw inspiration from work investigating how communities engage in {institutional design}, devising incentive systems that guide human behavior \citep{north_institutions_1991}. Prior work both draws on concepts from institutional theory to recommend new layers of governing to design for \citep{frey_thisplace_2019,schneider_modularpolitics_2021} and builds tools that would enable users to engage in decision-making about governance \citep{wang_pikaempowering_2024,zhang_policykitbuilding_2020}. For example, 
\citet{zhang_policykitbuilding_2020} presents {\em PolicyKit}, a software infrastructure that allows members of a community to author and implement new governance procedures on their home systems over time. Although these studies do not focus on inter-community governance, they suggest developing extensions, tools, and interfaces might help communities thoughtfully shape interactions they have with other communities. Considering institutional design as an inspiration for designing inter-community governance mechanisms also suggests expanding the scope of design beyond the technical: social rules, policies and incentives, and technical tools all become intertwined levers that can be ``designed'' to affect and implement inter-community governance. 

\subsubsection{Building Community Capacities}
Work across HCI focused on how collective action informs our view of how to support coordination. 
Notions of ``social design'' from \citet{ledantec_designcollective_2016} --- which emphasizes designing for social action with communities --- and ``infrastructuring'' community from \citet{ehn_participationdesign_2008} both emphasize building community capacities to act and organize \citep[e.g.,][]{li_outsite_2018, salehi_weare_2015, irani_turkopticoninterrupting_2013}. Recent work has proposed affordances and tools that might enable collective and communal forms of governance more generally \citep{mahar_squadboxtool_2018, jahanbakhsh_leveragingstructured_2022}, such as leaning on peers to help filter email harassment in \textit{Squadbox} \citep{mahar_squadboxtool_2018}. 

Communities engaged in collective action usually have a common goal they organize around \citep[e.g.,][]{matias_goingdark_2016}. Inter-community governance, however, does not suggest such a goal: rather, because communities on social media can vary broadly \cite{chandrasekharan_internetshidden_2018,weld_whatmakes_2021}, they may differ in ways that lead to conflict \citep{butler_crosspurposescrossposting_2011, eagar_whenonline_2015, datta_extractingintercommunity_2019}. Empirical work suggests direct conflict between communities is relatively rare, and driven by a small set of communities \citep{kumar_communityinteraction_2018}. At the same time, mechanisms that afford interactions across communities (e.g., cross-posting, multi-community membership, etc.) can make community differences more salient by placing content with incongruent norms side-by-side. When cross-community interactions are frequent, \citet{hwang_trustfriction_2025} note that communities may experience {\em governance frictions}: incompatibilities in community governance decisions such that the decisions of one community undermine the decisions of another. In some systems, communities address frictions with community-level blocks \citep{zhang_understandingcommunitylevel_2025}, but blocking can be a blunt tool for the many ways and reasons frictions arise. For example, two communities may use different filtering tools that lead them to moderate differently even if they have the same rules or values. Governance frictions articulate a critical design problem when strengthening community-level decision-making in social computing systems: although communities of users are afforded more voice and power --- which can enable nuance and context in decision-making \citep{chandrasekharan_internetshidden_2018} --- they are faced with new questions of how to negotiate their relationship with other communities in order to sustain their own goals, needs, and values. 
We aim to understand how social design can help communities navigate heterogeneity among themselves, beyond blocking. 

\section{Empirical Setting} \label{sec:setting} 
We turn to the case of the Fediverse, a decentralized social media network where many independent {\em servers} (i.e., {\em instances}) ``federate'' (i.e., agree to interact) through a shared protocol called ActivityPub. This protocol allows servers' respective users to exchange content and interactions from their home servers. A server can run on various types of software, including Lemmy (a community aggregator similar to Reddit), Pixelfed (an image-sharing platform), and PeerTube (a decentralized video service). The most widely used is Mastodon, which enables microblogging services similar to Twitter ($\mathbb{X}$) or Bluesky.\footnote{Bluesky is also a decentralized, protocol-based system. As it operates on a different protocol (AT Protocol) and bridging between the two protocols remains controversial, we do not include it in the scope of this work.} On the Fediverse, any individual can set up a server, although the skills and costs required to do so mean many join existing ones instead. Servers are usually managed by an admin, who holds the power to make decisions about data, moderation, federation, etc. In line with the definition of `community' presented in \S \ref{sec:bg_value}, we refer to servers and communities interchangeably. 

Although it has been around since the 2010s, the Fediverse saw a surge of growth in late 2022 \citep{jeong_exploringplatform_2024,cohn_fediversecould_2022,lunden_howmastodon_2022}, following concerns about Twitter after it was acquired by Elon Musk. The apparent autonomy of communities to make key social, organizational, and technical decisions remains a major draw. No centralized authority makes unbidden decisions for communities. Moreover, communities can choose who they interact with: federation is not obligatory and servers can `defederate,' usually to protect their community members from harassment or harm  \citep{theophilos_closingdoor_2024,melder_blocklistboundary_2025}. This ability is important as the Fediverse is notable as an space built by and for marginalized groups, particularly LGBTQ+ communities who were heavily involved in developing the protocol that the Fediverse operates on~\citep{karppi_ifnot_2023,gehl_activitypubnonstandard_2023}. Indeed, many communities use shared blocklists to defederate from known bad actors {\em en masse} \citep{zhang_understandingcommunitylevel_2025}. 

Two prominent examples illustrate this dynamic. In 2019, the alt-right platform Gab forked Mastodon’s open-source code and attempted to join the Fediverse, prompting mass defederation by progressive and queer-friendly communities \citep{zulli_rethinkingsocial_2020,robertson_howbiggest_2019}. In 2023, the announcement that Meta’s Threads would federate with ActivityPub led to the formation of the Anti-Meta Fedipact, a pledge circulated among server admins committing to block any Meta-owned servers that might appear on the Fediverse \citep{theophilos_closingdoor_2024}. 

While inter-community governance is relevant for any context with multiple communities, the Fediverse presents an ideal setting to explore the concept for two main reasons. First, communities are the main level of decision-making in this context. Although the software a community uses (e.g., Mastodon) can constrain it, communities are largely autonomous in that no central entity prescribes norms or practices to them and they can choose the suite of technologies they rely on. Thus, no centralized entity might step in to address frictions for communities. 
Second, because users across communities frequently interact, governance frictions regularly become apparent around incidents such as norm and privacy violations \citep{hwang_trustfriction_2025}. Although the protocol provides one sanctioning mechanism --- defederation --- it provides no mechanisms for communities to proactively identify frictions, resolve disputes, or coordinate actions. 
Both of these make the importance of designing for inter-community governance particularly critical in this context. 

\section{Methods}

\subsection{Design Workshops and Follow-up Interviews} \label{sec:methods_data}

\begin{table*}[t]
\centering
\begin{tabular}{l|llllll}

\textbf{W\#} &\textbf{P\#} & \textbf{Age}  & \textbf{Gender} & \textbf{Country} & \textbf{Main Fediverse Role(s)} & \textbf{Occupation} \\
\hline

\rowcolor{LightCyan}
\#1 & P1 & 55-64 & NB & USA & Admin, Writer & Software Engineer \\ 
\#1 & P2 & 18-24 & M & USA & Admin & Grad Student \\ 
\rowcolor{LightCyan}
\#1 & P3 & 25-34 & M & USA & User, Researcher & Operations Manager \\ 
\rowcolor{LightCyan}
\#1 & P4 & 55-64 & M & USA & Moderator, Organizer & Co-op Organizer \& Educator \\ 
\hline
\rowcolor{LightCyan}
\#2 & P5 & 45-54 & M & USA & Admin, Organizer & Non-profit Management \\ 
\rowcolor{LightCyan}
\#2 & P6 & 35-44 & M & Italy & Developer & Coder \\ 
\rowcolor{LightCyan}
\#2 & P7 & 25-34, & F & UK & Organizer & Head of Social \& Community \\ 
\#2 & P8 & 35-44 & NB & Spain & Moderator, Organizer & Freelance Artist \& Educator \\ 
\#2 & P9 & 45-54 & M & USA & Admin & Writer \\ 
\#2 & P10 & 35-44 & M & Canada & Moderator, Developer & Web Developer \\

\hline
\rowcolor{LightCyan}
\#3 & P11 & 35-44 & M & USA & Admin, Researcher, Developer & Computer Programmer \\ 
\rowcolor{LightCyan}
\#3 & P12 & 25-34 & M & UK & Organizer & Foundation Ambassador \\ 
\#3 & P13 & 35-44 & F & Netherlands & Admin, Organizer & OSS Trust \& Safety \\ 
\rowcolor{LightCyan}
\#3 & P14 & 35-44 & M & Netherlands & Writer & Writer \\ 
\rowcolor{LightCyan}
\#3 & P15 & 35-44 & NB & UK & Developer & Software Engineer \\ 
\hline
\rowcolor{LightCyan}
\#4 & P16 & 25-34 & M & USA & Admin & Cyber Security \\ 
\rowcolor{LightCyan}
\#4 & P17 & 35-44 & F & USA & Developer & Software Engineer \\ 
\#4 & P18 & 18-24 & M & USA & User, Researcher & Grad Student \\ 
\#4 & P19 & 25-34 & M & Czechia & Admin, Developer & Software Architect \\ 
\rowcolor{LightCyan}
\#4 & P20 & 35-44 & NB & USA & Admin & Software Engineer \\ 
\#4 & P21 & 55-64 & M & Argentina & Moderator, Organizer & Researcher \& Designer  \\ 
\#4 & P22 & 25-34 & M & Brazil & Moderator & Web Developer \\ 
\#4 & P23 & 45-54 & F & USA & Moderator & Student \\ 
\#4 & P24 & 45-54 & M & USA & User, Developer & Software Engineer \\ 

\end{tabular}
\caption{Participants across the workshop. Highlighted indicates follow-up interview.}
\Description{The table provides description of the participants, by columns: Workshop number, Participant ID, Age (in brackets), gender, (M, F, or NB), Country, Main Fediverse Role(s), and Occupation. In Workshop #1: P1 (Age 55–64, Non-binary, USA): Roles – Admin, Writer; Occupation – Software Engineer. P2 (Age 18–24, Male, USA): Roles – Admin; Occupation – Graduate Student. P3 (Age 25–34, Male, USA): Roles – User, Researcher; Occupation – Operations Manager. P4 (Age 55–64, Male, USA): Roles – Moderator, Organizer; Occupation – Co-op Organizer & Educator. In Workshop #2: P5 (Age 45–54, Male, USA): Roles – Admin, Organizer; Occupation – Non-profit Management. P6 (Age 35–44, Male, Italy): Roles – Developer; Occupation – Coder. P7 (Age 25–34, Female, UK): Roles – Organizer; Occupation – Head of Social & Community. P8 (Age 35–44, Non-binary, Spain): Roles – Moderator, Organizer; Occupation – Freelance Artist & Educator. P9 (Age 45–54, Male, USA): Roles – Admin; Occupation – Writer. P10 (Age 35–44, Male, Canada): Roles – Moderator, Developer; Occupation – Web Developer. In Workshop #3: P11 (Age 35–44, Male, USA): Roles – Admin, Researcher, Developer; Occupation – Computer Programmer. P12 (Age 25–34, Male, UK): Roles – Organizer; Occupation – Foundation Ambassador. P13 (Age 35–44, Female, Netherlands): Roles – Admin, Organizer; Occupation – OSS Trust & Safety. P14 (Age 35–44, Male, Netherlands): Roles – Writer; Occupation – Writer. P15 (Age 35–44, Non-binary, UK): Roles – Developer; Occupation – Software Engineer. In Workshop #4: P16 (Age 25–34, Male, USA): Roles – Admin; Occupation – Cyber Security. P17 (Age 35–44, Female, USA): Roles – Developer; Occupation – Software Engineer. P18 (Age 18–24, Male, USA): Roles – User, Researcher; Occupation – Graduate Student. P19 (Age 25–34, Male, Czechia): Roles – Admin, Developer; Occupation – Software Architect. P20 (Age 35–44, Non-binary, USA): Roles – Admin; Occupation – Software Engineer. P21 (Age 55–64, Male, Argentina): Roles – Moderator, Organizer; Occupation – Researcher & Designer. P22 (Age 25–34, Male, Brazil): Roles – Moderator; Occupation – Web Developer. P23 (Age 45–54, Female, USA): Roles – Moderator; Occupation – Student. P24 (Age 45–54, Male, USA): Roles – User, Developer; Occupation – Software Engineer.}

\label{tbl:pool}
\end{table*}

\begin{figure*}
    \centering
    \fbox{\includegraphics[width=0.865\linewidth]{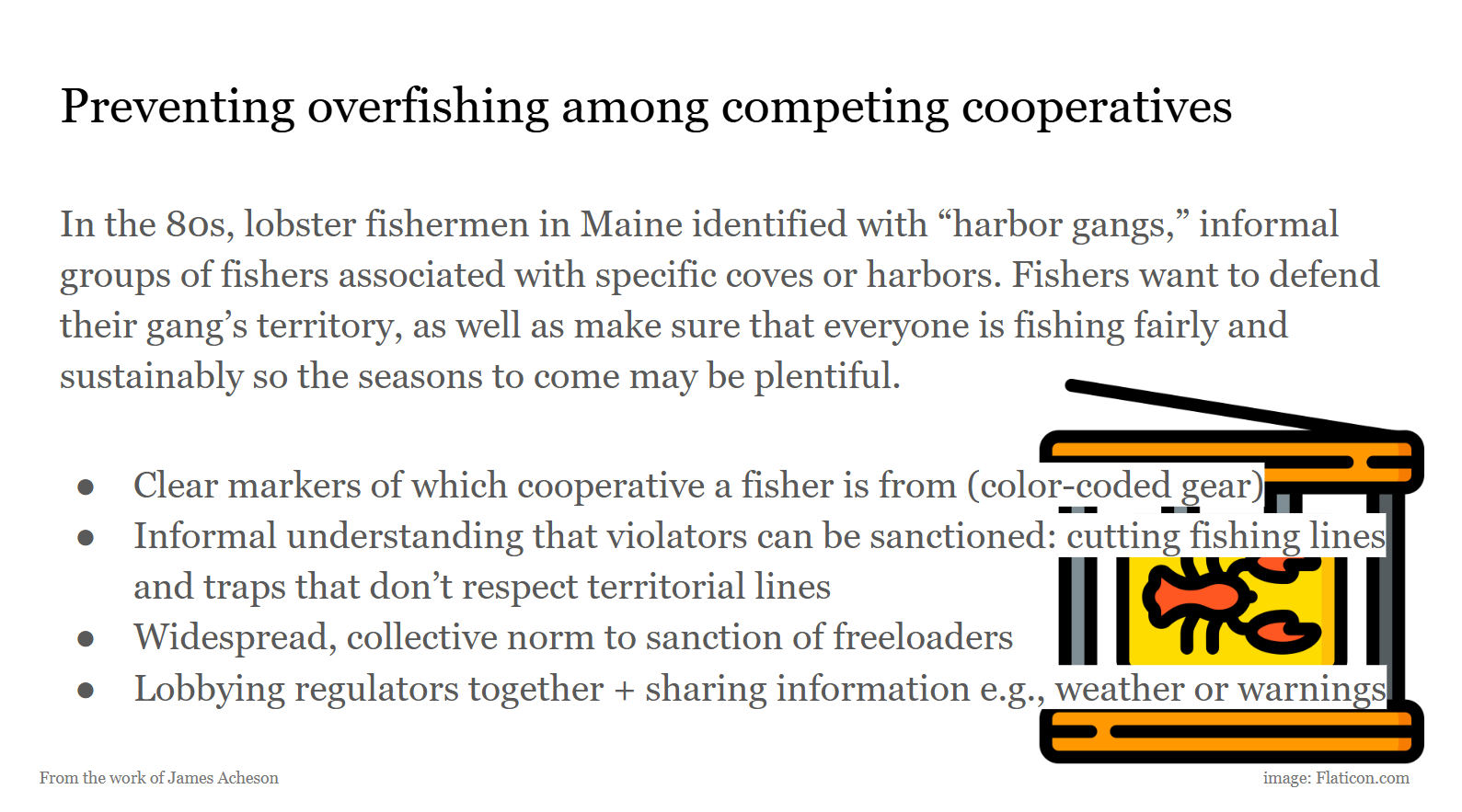}}
    \caption{A screenshot of the slide shown to participants to present the warm-up example of lobster fisherman, which the workshop facilitator talked through.}
    \Description{The figure shows a screenshot of a slide from the workshop, which describes an example that served as the warm-up activity. The text reads: In the 80s, lobster fishermen in Maine identified with “harbor gangs,” informal groups of fishers associated with specific coves or harbors. Fishers want to defend their gang’s territory, as well as make sure that everyone is fishing fairly and sustainably so the seasons to come may be plentiful. In bullet points: Clear markers of which cooperative a fisher is from (color-coded gear); Informal understanding that violators can be sanctioned: cutting fishing lines and traps that don’t respect territorial lines; Widespread, collective norm to sanction of freeloaders; Lobbying regulators together + sharing information e.g., weather or warnings.}
    \label{fig:lobsters}
\end{figure*}

We conducted four design workshops (June-July 2025), each running two to three hours over Zoom with small groups of individuals in the Fediverse. Discussions were scaffolded with a collaborative virtual whiteboard on Miro. Participants were offered an optional \$60 via Zelle as thanks for their time. The study was reviewed and approved by [anonymized university ethics review board(s)]. 

Participants were assigned to workshops primarily based on availability and timezone. A full table of participants (24 overall) can be found in Table \ref{tbl:pool}. Participant ages ranged from 18 to 64, and the pool included 16 men, four women, and four non-binary individuals based in ten different countries. All participants are active and deeply involved with the Fediverse and its governance, including those who are currently admins, moderators, developers, and organizers. Many of them had occupational backgrounds related to technology (e.g., software engineer, web developer, cyber security), reflecting broader patterns on the Fediverse. 

Following introductions, each workshop began with an example of cooperation among competing groups of lobster fishermen in Maine, inspired by the ethnographic work of \citet{acheson_lobstergangs_1988}. This example --- which describes how fishermen manage commons-based resources together for long-term, economic viability of their livelihoods as a whole (Figure \ref{fig:lobsters}) --- served as a prompt and warm-up activity for participants \cite{harrington_elicitingtech_2021,andersen_magicmachine_2019}. The focus on sustainability and collective benefit in the example oriented the tone of what values guided the workshop session. 
Participants were asked to share wonderings and reflections about the case, particularly in comparison to the Fediverse. For example, participants noted how the local, in-person nature of the case often created social norms and expectations that incentivized cooperation and made sanctions effective. 

\begin{figure*}[t]
    \centering
    \includegraphics[width=0.976\textwidth]{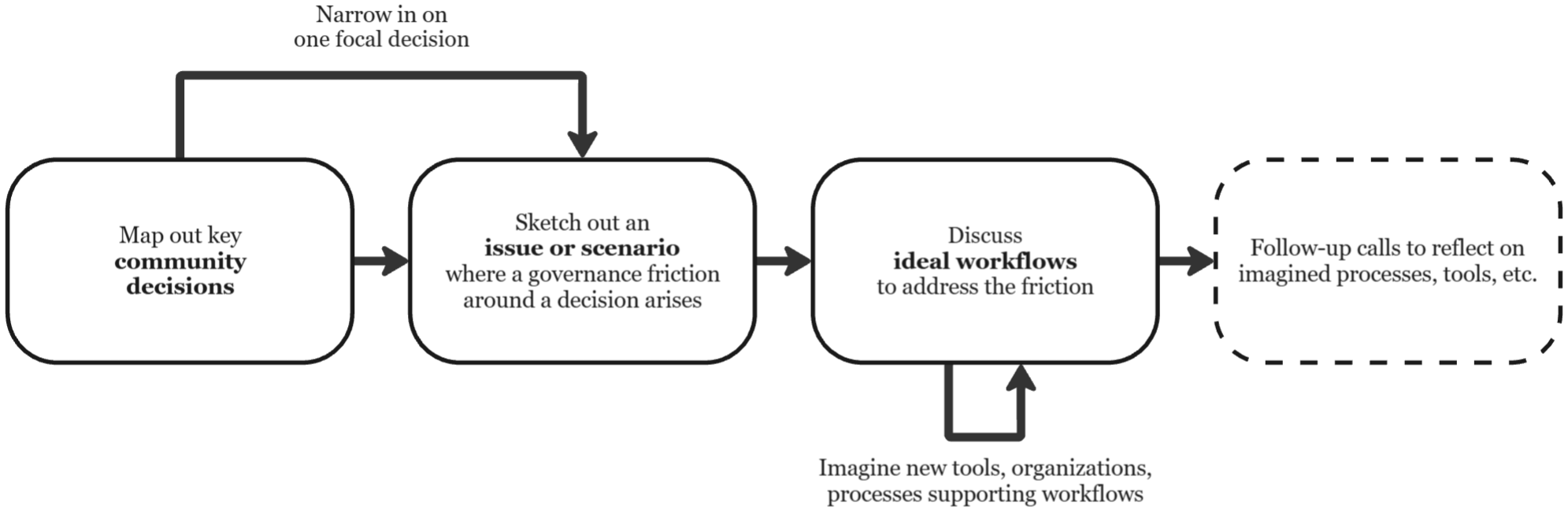}
    \caption{Guided flow of discussion during each workshop across two main break-out group sessions, followed by opt-in follow-up interview calls with individuals.}
    \label{fig:workshop_activity}
    \Description{The figure shows a flow of the main workshop activities: (1) Map out key community decisions, which participants then narrow in on one focal decision; (2) Sketch out an issue or scenario where a governance friction around a decision arises; (3) Discuss ideal workflows to address the friction, imagining new tools, organizations, and processes to support these workflows. Finally, there are optional follow-up calls to reflection on workshop discussion.}
\end{figure*}

The workshop then moved into its main activities, the flow of which is captured in Figure \ref{fig:workshop_activity} (with details in Appendix \ref{workshop_protocol}). Participants were first asked to collaboratively map out key community decisions, before selecting 1-3 that they found the most important. This was open-ended and we did not provide specific examples, encouraging participants to spitball. Then, participants took five minutes to sketch out a scenario where a {\em governance friction} \citep{hwang_trustfriction_2025} around any of the decisions arises. We leaned on this concept because it evokes the idea of design frictions, which typically refers to difficulties in user experience. Insofar that differences in how communities govern can cause challenges for one or the other, such differences cause friction. However, as \citet{cox_designfrictions_2016} notes, frictions can also prompt more mindful interaction. Likewise, we approach governance frictions as a generative opportunity.

Participants were then asked to brainstorm {\em ideal} ways to address the friction they had identified, ideating {\em potential} new tools that support them.
Participants were told tools need not be {\em technical} in nature: they could be resources, organizations, processes, and so on. 
Here, we drew inspiration from speculative design approaches \cite{dunne_speculativeeverything_2013,coulton_designfiction_2017,wong_elicitingvalues_2017}, which emphasize that people can imagine possible worlds by not only coming up with new technologies, but also by imagining alternative social values, organizations, rules, policies, and infrastructures. We found this perspective to be useful when considering the design of governance systems, which often require thinking about design at an institutional level, beyond technical tools alone. 

Across the activities, we served primarily as note-takers on the Miro, checking in with participants that notes captured were accurate. If conversation lulled, we also prompted participants for their thoughts to restart discussion. Finally, we conducted 13 optional follow-up interviews in July 2025 (this did not lead to more compensation); participants who completed these are indicated by the rows in blue in Table \ref{tbl:pool}. Interviews ran between 25-56 minutes each and were recorded, with the exception of one (P6). These calls enabled us to ask participants their reflections on the ideas discussed in the workshop sessions. Some participants were also involved in ongoing projects relevant to the issues discussed in workshops; conversations with these individuals also focused on the motivations and goals of their projects.

\subsection{Analytical Approach} \label{sec:methods_analysis}
The workshops yielded three main sources of data: the four virtual boards from the four sessions, the workshop recordings, and the follow-up interview recordings. Workshop and follow-up interviews were transcribed by an automated service, Otter.ai, and then manually corrected by members of the research team.

The first author began by reviewing the boards,\footnote{A screenshot of one board is shown in the Appendix as an example.} synthesizing across the sessions. We compiled a list of community decisions and ideas for social, organizational, and technical tools that participants discussed. Community decisions raised in the workshops, shown in Table \ref{tbl:decisions} in Appendix \ref{workshop_summary}, were clustered into groupings that might reflect broader dimensions of community governance. Tools were grouped based on similarity of ideas (e.g., some ideas were recurrent or highly similar) and are shown in Table \ref{tbl:tool_ideas} in Appendix \ref{workshop_summary}. These syntheses provided a rough map of topics and ideas across all participants, discussed in a team-wide meeting in July 2025.

We followed \citet{braun_usingthematic_2006}'s thematic analysis approach to identify challenges in inter-community governance. The first author conducted inductive line-by-line coding of the workshop transcripts, with an eye towards identifying motivations and concerns the tools responded to as well as surfaced. Moving between the transcripts and the synthesized mapping of ideas, the first author iteratively grouped the tools into core challenges. Four authors directly involved with data collection and cleaning met to discuss the stability of these groupings, agreeing on five major themes that felt both substantial and distinct. Then, the first author inductively coded the follow-up interviews, stress-testing the set of themes. This resulted in the development of one final theme that had not been captured previously. 

\section{Findings}
We organize our findings around six distinct challenges for inter-community governance. The first three sections focus on issues around information-seeking about and from other communities; the next two, on how relationships between communities are subsequently managed; and the final section, on questions around how interventions can be effective.

\subsection{Making Decisions About Community Governance Visible} \label{sec:findings_salience}

Participants across workshops noted that governance decisions (e.g., Table \ref{tbl:decisions}) by communities were often invisible to people outside of the community's admin and mod team, making it difficult to anticipate differences --- and any frictions that arose. 
They thus frequently proposed mechanisms that might make a community's governance decisions more salient. For example, in Workshop 4, P17 proposed a {community governance nutrition facts label}, inspired by existing tools for selecting open source licenses.\footnote{\texttt{https://choosealicense.com/}} P17 imagined such a label might describe decisions a community made along distinct dimensions of community governance --- its moderation process, a link to its rules, its decision-making model (e.g., coop, single admin, etc.), its goals for connectivity, and so on. Consistent structure would make it easier for people to assess a community, providing ``{\em much stronger signals}'' (P5) by enabling direct (and possibly automated) comparisons across a network of communities.

Participants suggested that consistent categories provide a ``{\em shared language/vocabulary}'' (P5, P10, P11, P12, P14) that \textbf{offer explanation about a community's behavior toward reducing tensions about frictions}. A lack of context about the basic goals of a community, or the size of its admin team, could lead to questions about ``{\em why the content is moderated the way it is [...] and why they’re seeing it in that space},'' producing ``{\em a lot of divide}'' (P17). 

Making dimensions of governance salient to others could also become {a guide for running one's own community}. 
As P9 remarked, this was crucial because each community impacts others: ``{\em you're not setting up just your little corner of the web. You are actually joining this - you are proposing to make a node on this network. [...] Think about {\em why} your community exists. How can you reflect your community in these rules?}'' 
Participants speculated that surfacing dimensions of governance could prompt intentional community decisions, such as around federation or moderation. P21 provided an analogy: 
\begin{quote}
    {\em Imagine someone who knows nothing about cars, choosing a car, right? [...] They have different [amounts] of power in their motors that goes with consumption of fuel. Also, there is capacity, speed, whatever, right? If you expose those facets to someone and explain what it means, [like]: with more power, the car can move things faster, but also it consumes a lot of fossil fuel, for example, versus renewable electricity. So exposing these things would be great, because it helps not only choose, but also understand what may be important to choose.} 
\end{quote}

\noindent P15 argued that ultimately, taking the time to reflect on these dimensions of governance at the start of a community's life could result in ``{\em less work in terms of governance and moderation, basically.}'' By making choices ahead of time, individuals could avoid ``{\em having to constantly make decisions about like: was this a problem? Is this person okay to participate?}'' 
At the same time, participants were concerned that such a tool could prescriptively embed values about what constituted good community governance, by explicitly laying out what one should make decisions about or possibly by creating default settings. P17 asked: ``{\em Do you even trust the creator of that tool? And how do you make that decision?}'' In order to instill trust, participants thus suggested developing such tools through consensus-based processes. 

\subsection{Collating Nuanced Information About Issues} \label{sec:findings_nuance}
Across workshops, participants wanted to know what issues other communities were dealing with, or had dealt with in the past, to anticipate real-time problems and get advice from others. This was often discussed in context of ideas such as a {moderator bulletin} or {shared modbox} (moderator inbox), envisioned to look across many communities to ``{\em pull together [public] information and make it more digestible}'' (P1) so that moderators could ``{\em handle the flood of crap}'' (P20).

As participants noted, some conceptually-similar spaces and resources already exist. Participants such as P1, P10, and P16 referenced IFTAS Connect, a forum for moderators maintained by a non-profit (IFTAS, or Independent Federated Trust and Safety).\footnote{\texttt{https://about.iftas.org/}} Other participants pointed to more private groups of moderators using Discord, email listservs, and other communication channels. A widespread existing resource mentioned across workshops was the \#Fediblock hashtag, which had been started by a group of queer femmes \citep{kiam_blacknessfediverse_2023} to help moderators across communities alert one another about bad actors. Although these resources were described as valuable, the information they provided was usually not systematic or contextualized enough unless one was ``{\em following this stuff extremely closely}'' (P1). For example, an individual might post about a community they had a bad experience with on the \#Fediblock hashtag, but not include evidence or an explanation.

Participants thus wanted bulletins and inboxes to \textbf{automatically present relevant, structured, and nuanced information about issues}: which ones were important, what were their main facts, how others were approaching them. 
Participants proposed pooling information about incidents, possibly through automatically shared moderation logs that were processed into discrete issues. Issues then might be algorithmically presented in order of ``trending'' problems or relevance for their community. Participants imagined being able to get deeper context about specific cases, reports, and content, sometimes through corresponding discussion threads where they could ``{\em sort of share and bounce ideas and like, how would you approach this?'}' (P10). P4 brainstormed: 
\begin{quote}
{\em Here's the hot issues coming up. For [an] issue, here's the receipts that I've seen, here's the context that I've seen. You know? Then people add in --- ``Oh no, there's this other important discussion here. Oh, here's how person A is thinking about it''. So ... A loosely structured, almost a separate timeline with more pooled information about: What's important? What should I know about it? And what are other people thinking about it in my circle?}
\end{quote}

\noindent Such a tool was imagined to help proactively anticipate problems in the network, as well as help community leaders learn from another. Participants also imagined having information through these spaces could help them make well-grounded decisions. For example, participants described wanting to distinguish between harm done by an individual vs a whole community: ``{\em Ideally, situations like these can be handled, like, pretty surgically, where whoever the individual bad actors are, they [individually] can just get removed}'' (P2). Being able to find enough context about a case would enable communities to make fair, nuanced judgments of who to hold accountable. For example, rather than banning an entire community, one might message its leaders to note that some of their rules are too broad and have allowed bad actors to slip by. 

\subsection{Managing What Information is Shared Across Trust Bubbles} \label{sec:findings_translucence}
Discussions also raised concerns about being too public or transparent. For example, preventing open access to discussions on bulletins can help protect involved individuals and communities. Participants also noted that communities could become vulnerable by making their decisions too visible, e.g., becoming targets for doxxing (P17) or making it easy to subvert blocklists (P16). 
Ultimately, however, P15 argued that not sharing information was not an option: ``{\em Security through obscurity [...] might work for a time, but [it] actually doesn't make you safe.}'' 

In all workshops, participants brainstormed how {information might be shared in tiered ways} through concepts, most explicitly named in Workshop 3 as {trust bubbles}: smaller networks or coalitions of other communities that they felt like they could trust. In sketching out the bulletins, modboxes, and other imagined interfaces for information-sharing, participants emphasized that these spaces did not need to contain everyone in the network. A community might also have multiple trust bubbles that they moved across, with some having more trust --- and more access to information --- than others depending on the nature of governance frictions they had. P6 put it succinctly: ``{\em I would really like that an admin [has] a trusted coalition that they can work with in total trust, or across different levels of trust --- because you don't have to share everything and all the details, but you can have different levels of trust where you talk, discuss [...] to try to address issues from a multifaceted point of view.}''

Per participants, trust bubbles already naturally exist in practice, with people reaching out to other communities that seem reliable (P5 noted that some software for running servers other than Mastodon, such as Misskey, seemed to support the idea). As such, it was ``{\em really the infrastructure that's missing for a trust bubble}'' (P14). Ideas such as bulletins and modboxes might, then, \textbf{enable the creation of subgroups where communities might share information in tiered ways across them}. For example, a subgroup of communities one finds generally reasonable might only get obfuscated or translucent insight about a community's decisions and processes; a closer, more trusted set of communities might get full access to discussions and unfiltered decisions. 

Discussions in Workshops 1 and 4 in particular touched on how sharing with more vs less trusted bubbles of communities would work, possibly through {abstract signals} that could still give people information about problems and other communities. 
P3 suggested ``some automated tool where maybe [it checks if] conversations have a different structure'' that might indicate reason for concern, examining ``the shape of data and the shape of conversations'', but not retaining their content. P16 spit-balled on more straightforward metrics that could be proxies, such as time to resolution for reports.
Signals needed to give individuals a sense of the {\em relevance} of a potential problem and the {\em urgency} with which they should act. They also wanted to be able to share signals of their own decisions or ongoing efforts to prevent misunderstandings or unnecessary conflict if frictions emerged. P4, for example, explained that because their community valued democratic decision-making, they were often slower in dealing with cases: ``{\em Our [server] could get blocked just because we have a longer process [...] And it would be great if we could communicate that [we're dealing with it] to the whoever it was that reported the issue.}'' While one might be able to just be fully transparent with others in very close trust bubbles, abstract signals could both protect the privacy of community decisions and help communicate about them. 

\subsection{Enabling New Relationships Between Communities} \label{sec:findings_accountability}
Because {communities are independently run, have different goals, and are not consistently nor equally impacted by incidents}, participants stated that ``{\em it can be very easy [for people] to say: {\em My server is good, and we're all good here. So, you know, it's fine [to not cooperate]}}'' (P11). However, some harms could only be addressed through cooperation, as noted by P24: 

\begin{quote}
{\em When you have someone who's jumping between different instances [and] changing their name or whatever --- like coordinated malicious activity --- then that's [...] where we have to talk with each other and come up with social norms across instances, or at least, consider that. [This] is its own special challenge.}
\end{quote}

To incentivize communities toward cooperation, participants suggested social mechanisms that might \textbf{make communities feel accountable to one another}, possibly inspired by the opening example. 
As noted in \S \ref{sec:methods_data}, each workshop had opened with a discussion about cooperation among fishers to prevent overfishing of lobsters in Maine. Participants pointed out how the shared economic interests of the fishers who lived and worked in the same region, compared to the diverse interests on the Fediverse, drew them to be responsible to one another.
In turn, participants considered how relationships among a set of communities --- i.e., a trust bubble --- might be configured to produce a similar effect: ``
{\em [The bubble] agree[s] to a minimal amount of shared rules and enforcement. And if you don't, then you might get removed from the [bubble's] list. [...] But then you could have alliances between alliances too, making it a true federation where it's not just either you're in or you're out}'' (P15).

The quote above suggests moving beyond a dichotomy of ``{\em either you're in or you're out}'' in how communities engage with each other. This was particularly fleshed out in the idea of {ladder of engagement}, raised by P16 in Workshop 4: ``{\em I start with a certain amount of trust --- or, like, start silenced --- and [...] if I want to federate with another instance, I would have to be like, okay, cool, I get this amount of federation by default, and then if I can prove to them [that I'm trustworthy], then I can get upgraded to platinum status.}'' The ladder of engagement might be a conceptual --- and encoded --- means through which the trust bubbles that participants discussed are formed, sustained, become tiered, and evolve. 

Participants hoped that such mechanisms \textbf{would enable more nuanced types of relationships between communities} beyond a federation-defederation dichotomy. P13 said:
\begin{quote}
    {\em One of the things that I would like to see for like this hypothetical modbox [...] is the ability to relay to each other, even if you need to block each other. Because let's say, you're not in a do-not-interact situation: it's not legal content, it's not harassment, stalking. It's just that our federation policies are in conflict, but not in a way that should cause a {\em trust} breakdown. Tooling should still support communication in that situation, somehow.}
\end{quote}

Participants imagined that a ``community of communities'' (P9) might eventually emerge, eventually leading to a {cross-community team of moderators} that could relieve each others' work. 
P23 explained how having a cross-community team would make it easier to ``{\em respond to things quickly and to chase down something and get it taken care of},'' ultimately having ``{\em a huge positive effect on the community, especially when they can see that they just report something and then it's taken care of.}'' In short, communities imagined governing together, sharing the burden of community governance through cooperation, support, and collaboration. 

\subsection{Customizing Community Governance Decisions} \label{sec:findings_interdependence}
As participants valued community autonomy, the shared nature of tools and resources raised concerns that the Fediverse would perceive them as centralizing or homogenizing forces. 
For example, the community governance nutrition fact labels (\S \ref{sec:findings_salience}) might be seen as demanding communities to conform. 
P9 humorously likened it to being a homeowner's association: ``{\em Lots of servers have rules that are like, {\em Everybody, be nice! Golden Rule. That's all you need.} And, okay --- how do we perhaps suggest that you want a little bit more of [substance in] that, without sounding like we're the homeowners association coming in and saying: {\em Nope, nope, your rules are all bad, you need to adopt ours instead}?}''

Participants thus emphasized the importance of first developing a consistent, shared language about community governance that could \textbf{prompt communities to make well-considered, independent decisions}. A few participants in Workshop 3 imagined this as being implemented through a {forkable, modular governance system} --- perhaps underpinning the generation of community governance nutrition facts --- where a community could search through a ``{\em user-friendly [...] set of options}'' (P12) containing decisions, practices, and processes that could be copied and tweaked from other communities. Participants envisioned that this system would cover different dimensions of governance (``{\em How do you structure a moderation team? How do you organize? What's the process when someone flags a post?}'' (P15), encoding them to be human and machine-readable. 
As communities went through the dimensions, P12 imagined that they would be prompted to reflect how a choice reflected its unique goals and needs: 
\begin{quote}
    {\em A community can ask itself: does it [...] want to federate? If yes, what kind of federation does it want? When it comes to building a code of conduct, which does it want? Like a super long, really in-depth code of conduct with examples of what is acceptable or not --- which has its pros --- or does it want a much more truncated code of conduct that is maybe a little bit more flexible and adaptable to different situations?}
\end{quote}

\noindent At the same time, one concern was that communities would simply copy without reflection. P15 even noted that having people ``{\em reinvent the wheel''} could actually be useful by leading communities {\em ``figure out a different way of doing things that either is better [...] or is just more appropriate for their needs or their particular community.}''

Crucially, participants did not envision the modular forking infrastructure to cover the entire network of decentralized social media at once. Multiple catalogs of governance decisions to fork from might independently exist, \textbf{reflecting the idea of many co-existing fediverses} that trust bubbles sat within. Participants frequently pushed back against the idea of one Fediverse, suggesting that this was a holdover from the image of Twitter/$\mathbb{X}$ as a global town square (P10 remarked: ``{\em There's never been such a thing! It's not even possible.}''). 
P6 pointed to an existing concept in the Fediverse, where communities existed in a ``archipelago'' structure: ``{\em forming an archipelago could be their way of creating their own Fediverse that could work well for them, for their use case, or for their communities.}'' As such, a community only draws on information or recommendations shared by different sets of trust bubbles in their part of the archipelago, depending on their goals and needs. The flexibility of who communities turned to appeared to relieve concerns about centralization: even if communities had shared tools and a shared language of governance, a plurality of governance practices was expected to emerge across them. In this sense, the ideas discussed do not smooth over governance frictions but enable communities to negotiate them.

\subsection{Minimizing Barriers to Inter-Community Governance} \label{sec:findings_adoption}
As noted earlier in \ref{sec:findings_nuance}, workshops often noted existing tools and resources like IFTASConnect or \#Fediblock that attempted to solve some of the problems raised in discussion. These projects had struggled to see widespread or consistent use in the decentralized network. Likewise, any attempt to support inter-community governance would face the challenge of adoption: ``{\em We can come up with 1000 good ideas, maybe quite easily, but getting, like, the admin of random.social to implement it for their 200 users is gonna be a huge, huge headache}'' (P12).

Participants suggested that the best way to overcome this was to make any technical or infrastructural tools \textbf{easily integrated or built-in to existing technologies} so that people could avoid ``{\em splitting [their] energy and focus}'' (P6). Across workshops, participants such as P4, P13, and P23 noted that they {already had too many tools that their community was juggling}. This made logging into and learning to use yet another tool for inter-community governance unappealing, especially as the need for inter-community governance is not always obvious in the daily work of moderators. At the same time, when questions or harms arise, the benefits of and need for inter-community governance become clear. One intuitive path forward might be to foster relationships with developers of Fediverse platforms used to run a server, to encourage them to build the kinds of tools participants described in workshops since such software ``{\em determines a lot of [what] different governance models become available to you}'' (P11). 

However, communities do not all use the same software and the available software on decentralized social media has evolved over time. Participants thus instead envisioned {building a software-agnostic layer}, as articulated by P11: ``{\em 
It's more like standard tools that can be integrated into the platform, rather than the platform comes up with its own bespoke tools, right?}'' This suggested building extensions, APIs, and other meta-level infrastructure that could be layered with existing software, akin to another protocol.

Having been told that a tool could include organizational and social approaches, participants also \textbf{questioned whether a ``tool'' had to be technical in nature}. 
At the core, any tools devised to support inter-community governance had to be ``{\em social tools}'' (P5). P4 stated: ``{\em I think that really, at a deeper level, this is about moderators across communities talking to each other and building enough of a community that we're able to really be in touch and know who to talk to and when to have these conversations [about frictions]}.'' Participants pointed out that simply having {regular open moderator calls} might help foster the community of communities they envisioned. P13 imagined how this simple sense of community might in turn help encourage the adoption of technical tools that made dealing with frictions easier: ``{\em Wouldn't it be nice if there was a tool [...] and I announced [it] to my trust bubble: {\em Hey, did you see this thing?} [...] And now, it bops, bops, bops [throughout the trust bubbles], and suddenly 50\% of the Fediverse is doing it, and the rest can do it by osmosis.}'' In short, a cultural shift might be facilitated by technologies --- but ultimately, does not require  them.

\section{Discussion}
\begin{figure*}[ht]
    \centering
    \includegraphics[width=0.95\textwidth]{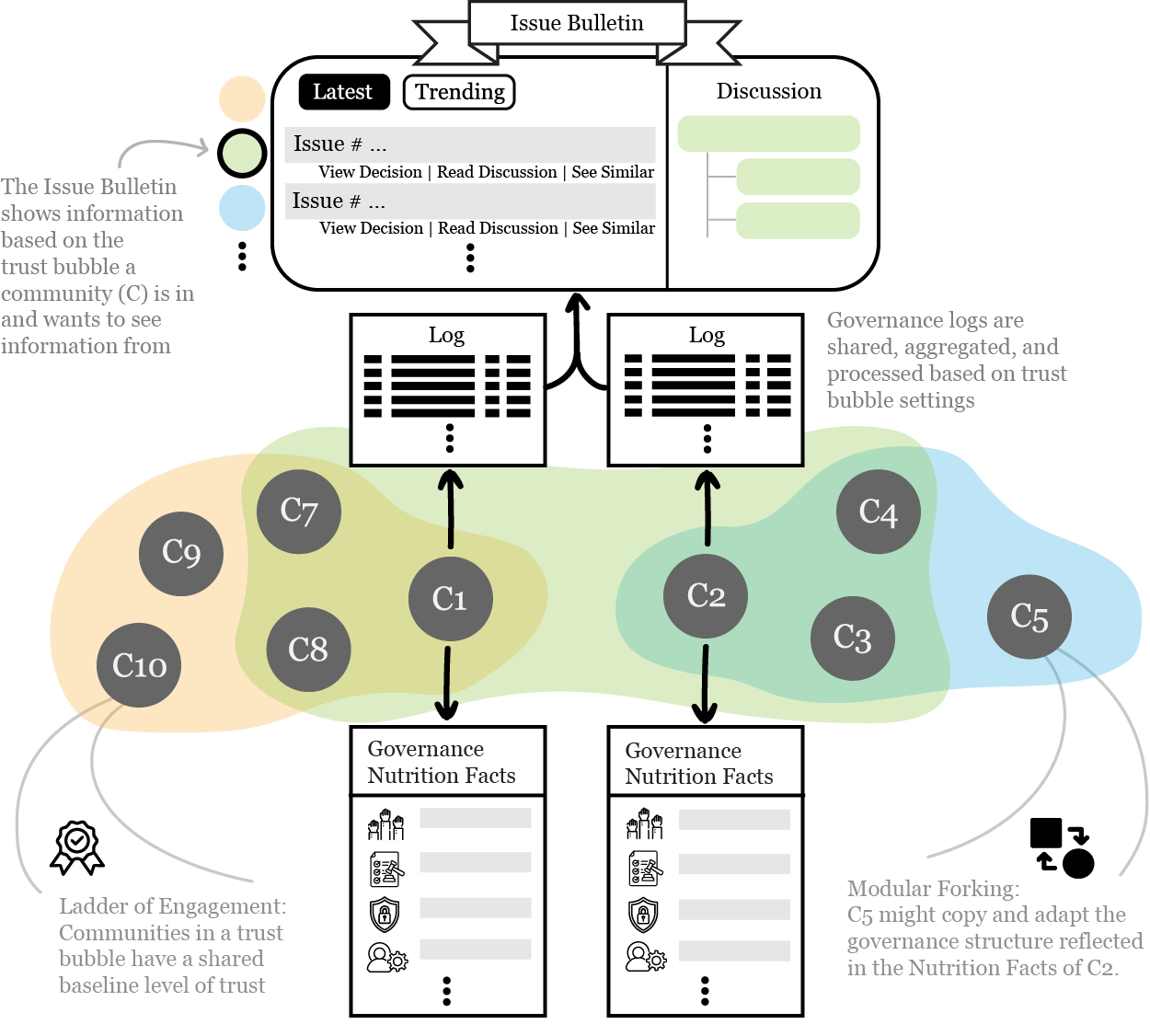}
    \caption{Visualization of the approaches imagined by participants as an ecosystem of shared resources, spaces, and automated tooling. A community (C) shows its ``Governance Nutrition Facts'' label, which can be generated through a user-friendly interface that modularizes dimensions of community governance. Communities can fork and customize these labels. Communities sit in polycentric ``trust bubbles,'' or coalitions of communities, that are indicated by the colored clouds grouping communities. These trust bubbles help structure how communities share and receive information about governance from one another, shown in an ``Issue Bulletin'' for each trust bubble. The Issue Bulletin automatically processes governance activity logs into a database a community can review. It also enables communities to engage in discussion.}
    \label{fig:lifeworld}
    \Description{The figure gives a conceptual overview of all the different ideas proposed by participants, in one ecosystem. It shows ten gray circles, each representing a community. Three colored blobs covering different but overlapping subsets of them indicate the `trust bubble' they are in. The figure shows how communities each have Governance Nutrition Facts, which looks like a label with a consistent icons on the left representing a dimension of community governance. Presumably, communities note their decision for each dimension. On the right of the labels, there is a note about modular forking, which enables communities to copy and adapt the Governance Nutrition Facts from one another. The image also shows a community exporting its activity log to an `Issue Bulletin,' a shared space for the community's trust bubble. The Bulletin has a split pane, where the left pane shows aggregated issues and meta-information about it and the right pane shows discussion threads. Notably, left of the Issue Bulletin are three dots that correspond to the trust bubble blobs. Communities can select which trust bubble's Issue Bulletin they want to see and contribute to. The figure also has a note about the `ladder of engagement,' which is a concept to ground how communities change their information sharing depending on where on the ladder they are with others. For example, communities that are mutually low on the ladder of engagement may only share abstract metadata. Communities that are mutually high on the ladder of engagement might send much more detailed information about their processes.}
\end{figure*}

\subsection{Modularity, Polycentricity, and Forkability: Principles for Inter-community Governance}
Our findings suggest a collection of resources, norms, and spaces that can enable communities to collaborate and cooperate on governance issues from the bottom-up. In Figure \ref{fig:lifeworld}, we sketch them as one ecosystem of inter-community governance on community-run social media. We present this illustration as a conceptual artifact that can prompt future design work aiming to support, shape, and scaffold dynamics of inter-community governance. 

Some of the concepts presented above could be easily developed in the near future. However, as our findings note, the success of any system suggested in Figure \ref{fig:lifeworld} is likely to require shifts in broader social expectations around how communities operate on the Fediverse. For example, many workshop participants envisioned decentralized social media as consisting of ``many fediverses,'' an idea that they stated was considered controversial by those who want the benefits of scale. The specific tools we describe are not technical solutions, but aim to support norms of communities engaging with one another on governance questions when needed. \textbf{We emphasize devising infrastructures that can create opportunities for communities to organize and reshape norms together}. For example, a ``nutrition label'' alone could help a community communicate its practices to existing and potential members. However, a consistent, structured labeling schema for it would allow communities to engage in inter-community governance much more easily. To illustrate: as a machine-readable system, a repository of nutrition labels from a wide swathe of communities could enable people to quickly flag and review potentially incompatible communities (or find compatible ones). 

Our perspective builds on prior work from HCI that intervenes in technical infrastructures that standardize processes (e.g., decision-making) or produce new possibilities (or constraints) \citep{edwards_infrastructureproblem_2010}, as well as examines human infrastructures that do work to coordinate and manage these systems \citep{lee_humaninfrastructure_2006}. Designing infrastructures presents opportunities for discussing social norms and discussing what types of decisions should be made at a global versus local level \cite{edwards_introductionagenda_2009}. As we note in \S \ref{sec:findings_adoption}, this may start as simply as hosting regular open calls for community leaders. Recent efforts such as FediForum,\footnote{\texttt{https://fediforum.org/}} a virtual convening of Fediverse users in the style of ``unconference'' sessions, are already undertaking such initiatives, reflecting the evolving nature of the Fediverse. 

Technical infrastructuring can enable the kind of interactions and relationships between communities that must be in place for routines of inter-community governance to emerge. We walk through the imagined ecosystem to highlight three interdependent principles that would enable future design work to meet the challenges of coordinating governance across communities: \textbf{modularity}, \textbf{polycentricity}, and \textbf{forkability}. 

\subsubsection{Modularity: The Capacity for Parsimonious Communication}

\begin{itemize}
\item[\ding{43}] The costs communities induce in communicating governance decisions came up repeatedly. These costs occur not necessarily because communities hide information from one another, but because the information shared about governance practices (e.g., structures, procedures, logs) by communities can be highly inconsistent --- not only in what is shared, but also in the language used. 
\end{itemize}

\noindent The envisioned ecosystem of inter-community governance in Figure \ref{fig:lifeworld} begins with multiple autonomous communities (C1, C2, C3, and so on), some --- but not all --- of which are federated together. Each community generates and publicly displays their ``Governance Nutrition Facts:'' a label summary of the governance structure of that community, including brief descriptions along consistent categorical dimensions of governance (such as decision-making process, rule set, data security, and the admin team). These dimensions, as well as the potential choices within each dimension in the label, are modularized to make comparison across communities meaningful. For example, under rule set, one community might indicate they follow the ``Mastodon Server Covenant''\footnote{\texttt{https://joinmastodon.org/covenant}} plus a small set of stricter rules related to their community's topical focus; another might only indicate that they follow the Covenant. 

A community creates their Governance Nutrition Facts label via an interface through which it can cherry-pick modular choices for each dimension of community governance. This presumes that communities have a shared language about important governance decisions and how they are made, even if communities may vary in their selection. A modular system thus \textbf{reduces the potential costs of communicating about community governance}. Displaying a consistent set of dimensions and potential choices, \textbf{modules allow for more parsimonious reading and review of community governance decisions} by humans compared to raw textual description. 
Modules can also make community governance \textbf{machine-readable}. This can \textbf{enable a new suite of automated tools around governance}, e.g.,  identifying communities with similar practices, or alerting communities of potential governance frictions. 

\subsubsection{Polycentricity: The Capacity for Overlapping, Nuanced Relationships}

\begin{itemize}
\item[\ding{43}] Workshops noted that communities would benefit by getting information from one another and possibly sharing their moderation work. However, communities neither trust nor want to exchange information with each other equally. A lack of infrastructure to coordinate with trusted others undermines the potential to share the burden of community governance.
\end{itemize}

\noindent Communities create or join trust bubbles: sets of communities that they indicate a level of trust in, even if they are not all federated. In Figure \ref{fig:lifeworld}, trust bubbles are represented by shaded blobs encompassing communities. 
Participants imagined that trust bubbles would help communities gather information about governance practices and issues. For example, communities could automatically share logs of their governance activity that are aggregated in the ``Issue Bulletin'' for the trust bubble. Crucially, the bulletin is not a global repository; each trust bubble can instantiate its own Bulletin, and communities might navigate multiple Bulletins according to the various trust bubbles they are in. How each Bulletin processes and displays information differs based on the trust bubble's trust levels, which can vary over time. As with the modular system for community governance labels, trust bubbles have a modular set of choices for how to share information as well as engage in discussion. Trust bubbles might set processes to edit these choices using existing tools such as {\em PolicyKit} by \citet{zhang_policykitbuilding_2020}. 

Trust bubbles envision a polycentric\footnote{For overview of polycentricity as a concept in the context of social media, see \citet{jhaver_decentralizingplatform_2023}.} system of evolving coalitions that a community is a part of: multiple, overlapping sets of communities that coordinate with one another in different ways, for different reasons.  
Polycentricity \textbf{enables a community to be in multiple trust bubbles} that may or may not overlap; it also suggests \textbf{multiple kinds of coordination} across those trust bubbles. For example, a community may be in a smaller trust bubble that actively shares raw moderation reports, and a broader one for tracking emerging issues (e.g., spam, scams, raids) on the network. In short, polycentricity involves \textbf{supporting nuanced and diverse governing relationships between communities} that evolve over time. 

This dynamic and relational aspect of polycentricity in trust bubbles distinguishes it from existing inter-community governance. For example, the AT protocol that underpins Bluesky affords ``composable moderation'' by enabling a community to label content (i.e., for filtering) independent of the app they use \citep{bluesky_composablemoderation_2023,fraser_socialmedias_2025}. Others can subscribe to these labels, allowing them to reduce their moderation burden. Trust bubbles imagine infrastructures that enable multiple communities to collaborate on such resources; different bubbles may design different solutions. The idea of the Issue Bulletin illustrates how this might manifest more concretely, as well as the potential benefits of \textbf{expanding how and when {communities} are able to support one another on governance problems}. 

\subsubsection{Forkability: The Capacity for Variation}

\begin{itemize}
\item[\ding{43}] Sharing information, resources, and tools raised concerns about whether communities could meaningfully sustain autonomy within an interdependent network. Although polycentricity can offset such pressures by allowing communities to exist in varying trust bubbles, participants emphasized needing to ensure communities could make decisions according to their own needs and goals. 
\end{itemize}

\noindent When a community generates their Governance Nutrition Facts label, they can fork a version from other communities and adapt each dimension of community governance. The options available in the forking system can draw from the Nutrition Facts of all communities in the network to provide a modular catalog of decisions to fork from along each dimension. It can also be filtered by trust bubble. In turn, communities can also contribute to this catalog, so that other communities might reference it in the future. The shared language that the modular forking system is based on is descriptive, not prescriptive. Along each dimension of community governance, the system prompts the community to reflect on whether the forked decision reflects their values and goals. 

Forkability reduces the labor it takes to craft governance processes. Communities already copy and customize governance practices based on what they see in other communities \citep{tosch_privacypolicies_2024,kiene_relationalorigins_2025}, but are prone to path dependencies. We suggest that forkability also requires a heterogeneous catalog of potential choices to copy from, as envisioned in the modular forking system in our ecosystem. Forking is not just about copying; it is about \textbf{adding, customizing, and discarding modular components as needed}. Forkability ensures a community can tailor their governance practices to their independent needs. Communities can add their forks as new modules. Forkability thus also emphasizes that \textbf{modular choices can evolve over time}. 

\subsection{Tensions in Implementing Inter-Community Governance} \label{sec:disc_implementing}
Modularity, polycentricity, and forkability are interdependent; the lack of one would undermine the others. Without modularity, communities lack the capacity for parsimonious communication that enables them to navigate nuanced relationships and distinguish themselves in forking. Without polycentricity, communities lack the capacity for diverse cooperative relationships that allow them to engage with one another, assess modular options, and know when to fork. Without forkability, communities lack the capacity to maintain their autonomy, sustain diversity, and evolve modular governance elements that fit their needs within a polycentric system. All three principles must be attended to when implementing infrastructures for inter-community governance in design. 

The principles, and their interdependence, also reveal tensions and tradeoffs in inter-community governance. The emphasis on {\em parsimonious} communication in modularity underscores that the opacity and inconsistency in community governance makes it labor intensive to assess the good faith and legitimacy of communities online.  
The emphasis on evolving relationships with {\em multiple} bubbles of communities in polycentricity responds to concerns about potential re-centralization if communities begin to coalesce and coordinate. Likewise, the emphasis on autonomy and variation in forkability aims to avoid \textit{de facto} re-centralization through institutional homogeneity. 

Below we reflect on the two key tensions around labor and recentralization, and their implications for future work.

\subsubsection{Labor}
Even in a speculative mode, participants expressed hesitation about adopting inter-community governance tools. Community leaders already struggled to juggle a patchwork of tools for governing their communities. Inter-community governance was not always a priority. It became salient when a harm occurred due to a governance friction (e.g., norm violation). Under the circumstances, on-boarding, logging onto, and participating in a new governance system could seem unappealing, just another dimension of governance labor for communities. For example, crafting a tailored community nutrition label requires care and time, even if there are useful default modules that make communication more parsimonious. 
Likewise, an Issue Bulletin could produce a new source of information overload. Polycentricity suggests that communities could have many Bulletins, compounding these problems.

Attributes of our participants---all deeply invested in the work of community-building---may have led them to prefer tools that entail care and participation. Recall that many communities end up copying things like policies \citep{tosch_privacypolicies_2024,kiene_identitylegitimacy_2024}, and face risks of burnout \citep{schopke-gonzalez_whyvolunteer_2024}. 
Our participants were aware of broader trends and explicitly raised questions of feasibility, adoption, and sustainability. Future research should attend to the barriers to participating in these infrastructures as they are implemented. In the context of corporate-run platforms, failing to do so can risk enabling platform operators to exploit the care and labor of communities \citep{li_ethicaltensions_2022,jhaver_personalizingcontent_2023,matias_civiclabor_2019}. 

To reduce potential barriers, we propose that infrastructures be implemented via extensions rather than new standalone apps. 
The specific design of the tools described in Figure \ref{fig:lifeworld} will shape the labor associated with inter-community governance. Our findings suggest that inter-community governance could be leveraged to \textit{reduce} the governance burdens communities face. This was most salient when discussing trust bubbles: participants imagined pooling moderation work with other communities. In a highly-trusted bubbles, communities might automatically share moderation decisions that are then applied by other communities in the bubble (e.g., deleting the same post, or flagging similar content), reducing the volume of things communities must independently deal with. 
Prior work in CSCW proposes similar ideas of leaning on assessments by trusted peers \citep{jahanbakhsh_leveragingstructured_2022, mahar_squadboxtool_2018}. As \citet{jahanbakhsh_leveragingstructured_2022} find in the case of misinformation, preferences about the exact degree of automation in such systems (e.g., algorithmic filtering) vary widely. Future work testing prototypes should focus on these questions of labor, evaluating how to relieve the burdens of use --- and toward strengthening the sustainability of communities.

\subsubsection{Re-centralization}
Communities may feel pressure to conform to ``good'' governance practices based on what they see in the Issue Bulletin, the trust bubbles they are a part of, or the community nutrition labels of large or influential communities. Participants raised concerns that trust bubbles could simply become spaces for re-centralization, or that nutrition labels could become prescriptive and lead to homogeneity. For example, making nutrition labels ``forkable'' (i.e., letting people copy and then customize from another community) could result in mindless copying or homogenization \citep{gaughan_introductionreadme_2025}. Given the concerns around labor noted above, such risks seem real . 

Communities often follow similar institutional patterns \citep{shaw_laboratoriesoligarchy_2014} and organizational structures \citep{schneider_adminsmods_2022,schneider_governablespaces_2024}, in addition to rules \citep{kiene_identitylegitimacy_2024}. Some similarities may reflect shared goals or well-known best practices \citep{kraut_buildingsuccessful_2012} --- not all communities can be wildly different experiments in self-governance, and communities can learn from one another \citep{kiene_relationalorigins_2025}. However, robust defaults can reduce barriers to entry, i.e., in creating a nutrition label that truly reflects a community's goals. For example, even if a community forks a nutrition label, explicitly presenting a variety of choices about decision-making models or data retention policies can facilitate tailoring. We propose developing an open standard toolkit of modular ``community governance choices'' that covers a range of potential community configurations. Future UX and UI work can investigate how such toolkits balance between community autonomy and learning or copying.

Regardless, with an inter-community governance ecosystem, increased opportunities to interact about governance may increase isomorphic pressures \citep{dimaggio_ironcage_1983,caplan_isomorphismalgorithms_2018} that push communities to make {\em non-optimal} decisions for themselves. 
Our participants suggested being able to move across {\em multiple} trust bubbles that hold distinct norms about what is shared, to whom. Thus, being `in' a bubble can mean different things, enabling communities to have diverse types of relationships with other communities. 
Future work defining a taxonomy of trust to help communities navigate a ``ladder of engagement'' and underpin concepts like trust bubbles can shape how communities preserve autonomy amidst homogenizing pressures.

\subsection{Beyond the Fediverse} \label{generalizability}
The protocols that underpin systems such as the Fediverse tend to specify broad affordances for inter-community interactions, but provide little-to-no ways to support inter-community dynamics, reduce frictions, or help community leaders solve typical problems. This makes the need for a layer of collective governance particularly pressing. However, interactions flow across communities in most social computing systems. We anticipate that other systems beyond federated, decentralized social media would benefit from inter-community governance.

Concepts like the community nutrition label or the trust bubbles evoke general questions about building and sustaining community. Thus, the ideas and design principles that underpin inter-community governance may be broadly applicable to a variety of social computing systems involving multiple communities, including lo-fi or indie ones. For example, a group starting a wiki for happenings in their city may want to connect with the leaders of other place-based communities to know how to deal with local emergencies, breaking news, and so on. Building the capacity to self-govern through ecosystems of inter-community governance can support communities to run their own systems, and we imagine that inter-community governance could support a diverse array of such systems. 

Even in centralized platforms, community-level decision-making makes inter-community governance salient \citep{seering_reconsideringselfmoderation_2020,jhaver_designingmultiple_2021,li_measuringmonetary_2022}. Communities in more centralized contexts already govern in relation to one another, and could benefit from the kind of information-sharing and coalition-building imagined by inter-community governance \citep{kiene_relationalorigins_2025}. Inter-community governance may thus inform platform design, such as by justifying more back-end capacity for community organizers using Facebook Groups or Slack workspaces to coordinate with one another. 
Future work can also learn from the existing community governance features encoded in widely-used platforms. For example, Reddit provides a side panel with important information about the community for the platform: individual admins, community bookmarks, and community ``flairs.'' Although this does not necessarily reflect inter-community governance practices, empirical assessment of what is already made visible about community governance in popular systems can provide guidance on what still needs to be articulated.

\section{Limitations}
Our findings are shaped by the subjectivity and self-presentation of our participants, as well as the interpretative nature of our qualitative work. We aimed to build a participant pool with a broad range of experiences on the Fediverse, but focused on individuals who were interested in governance questions for the purposes of our workshop. This allowed us to learn about perspectives of community organizers and developers, but future work evaluating the themes and ideas presented in this work with community members are likely to give insight into notions of fairness, transparency, and participation in the kinds of governance dynamics described in our findings. Crucially, as we noted in \ref{sec:disc_implementing}, the participants in our workshops are heavily invested in governance. They had not only thought about practices and processes extensively, but also were willing to invest extra time, labor, and care in nurturing online spaces in ways that many others---even community moderators---may not. Future work should consider other perspectives and explore how to manage the potential barriers and costs to this kind of coordination work, particularly as we see growing calls to give more control to communities. 
In \S \ref{sec:disc_implementing}, we noted that inter-community governance is more useful when it is implemented as an infrastructural extension, rather than when it is implemented as standalone apps that require new sign-ups or technical learning. 
User studies with prototypes operationalizing some of the ideas described by participants would help offer more specific recommendations for designing such a layer, as well as identifying barriers to use. Finally, testing such prototypes outside the Fediverse will be important to refine our understanding of how and when to implement inter-community governance in a diverse array of contexts. 

\section{Conclusion} \label{sec:conclusion}
Drawing from four workshop sessions with Fediverse organizers and developers, we make a case for inter-community governance in social computing systems, presenting a vision of concrete ideas and principles that may guide future design. Community-level decision-making plays a critical role in enacting governance in social computing systems, but faces persistent challenges and harms. Inter-community governance offers a `meta' approach to addressing these issues by envisioning an infrastructural layer that support interaction and relationships among communities, so that they may govern together while preserving autonomy. Focused on the case of community-run social media, discussions with participants underscore six challenges that inter-community governance grapples with: making community governance decisions visible; collating nuanced information about issues; managing what information shared, to whom; enabling new relationships between communities; being able to customize community governance decisions; and minimizing technical barriers to adoption. 

To meet these challenges, we suggest that future inter-community governance design efforts focus on three interdependent principles: modularity, forkability, and polycentricity. By synthesizing the ideas from workshops into an ecosystem of tools, interfaces, and mechanisms, we demonstrate one way these principles might operate together. The imagined ecosystem maps a path forward for future work to interrogate and experiment with inter-community governance, toward building infrastructure that supports communities in a variety of social computing systems. 

\begin{acks}
We are grateful to our participants who were generous with their time and thoughts in workshops, follow-up calls, and beyond. We thank our thoughtful reviewers for their constructive feedback on this piece. An in-progress version of this work was shared at a Siegel Research Fellow meeting as part of the first author's fellowship.
\end{acks}

\bibliographystyle{ACM-Reference-Format}
\bibliography{refs}

\appendix

\section{Workshop Protocol} \label{workshop_protocol}

The workshop began with a warm-up example of cooperation in commons-based resource management based on the work of \citet{acheson_lobstergangs_1988}, presented with the slide shown in Figure \ref{fig:lobsters}. We asked: \textit{Go around and share anything loose thoughts about what we noticed, wondered about, or found interesting with this example. What made cooperation possible, and what were the challenges they faced in their context?}

Next, we introduced the concept of \textit{governance frictions}, which we defined as ``{\em refer to incompatibilities in community governance decisions that enforce the goals of one community but undermine the goals of another}.'' The workshop then moved into a series of 20-30 minutes of breakout discussions, with the following written prompts:
\begin{enumerate}
    \item \textbf{Governance frictions seen and encountered in the wild}: \textit{What are some concrete, specific decisions made about how to run and govern a community on the Fediverse? These can be social, technical, organizational, etc.}
    \begin{itemize}
        \item \textit{Narrow down to 2-3 decisions you feel are the most pressing and important. Why? }
        \item \textit{For each decision: what choices might other communities make about that decision? What problems or challenges result, if at all? Who deals with them? }
    \end{itemize}
    \item \textbf{Ideal workflows}: \textit{Create a flow chart of steps in an indeal workflow for handling the friction your group has selected. You're in an \textbf{ideal} world: perhaps your workflow is part of new policies, technologies, agreements, infrastructures, etc. Break down into the steps and per step, decisiosn that need to be made.}
    \item \textbf{Tools to support our ideal workflow}: \textit{(1) Brainstorm a \textbf{speculative} tool that supports our ideal workflow in some way. (2) Tell us how it works, including: the \textbf{goal (+what step/s of the workflow it's a part of), features, limitations, who is using the tool, who makes decisions about it, any relevant policies or processes}. Remember we're in an ideal world - the tool might not be immediately possible in our world!}
\end{enumerate}

We verbally noted to participants that by focusing on \textit{ideal} workflows and \textit{speculative} tools, we aimed to imagine possible sociotechnical futures - not only focus on how governance currently works on the Fediverse. For Workshops 3 and 4, Activity 2 and 3 were merged into one activity, as we found that participants tended to talk about speculative tools immediately with workflows, and for time.

\section{Summary of Ideas from Workshops} \label{workshop_summary}
The following tables provide a succinct summary of core points of discussion in the workshops, particularly the kinds of decisions participants felt were important and concepts/potential tools or resources ideated on.

\begin{table*}[h!]
\centering
\begin{tabular}{|ll|}
\hline
\multirow{6}{*}{Technical and Infrastructural Decisions}
    & use of complementary platforms like Loomio \\
    & server and hosting infrastructure \\
    & what relays used \\
    & tools like bots \\
    & platform software \\
    & member-facing features \\
\hline
\multirow{9}{*}{Organizational Decisions}
    & number of admins \\
    & ownership \\
    & internal communication norms \\
    & formalization as an entity \\
    & funding model \\
    & legal obligations \\
    & governance capacity \\
    & decision-making model (e.g., cooperative) \\
    & rule enforcement process \\
\hline
\multirow{13}{*}{Normative Decisions}
    & codes of conducts, rules, and the concreteness of those rules \\
    & content warning requirements \\
    & terms of service \\
    & use of blocklists \\
    & documentation of community practices (for transparency) \\
    & membership (e.g., open vs closed registration) \\
    & onboarding practices \\
    & federation policy \\
    & bridging (to other services) \\
    & open to threads/meta, mastodon.social, and other at-scale services \\
    & relationship to third-party groups like IFTAS \\
    & privacy policy \\
    & data permanence and maintenance \\
\hline
\end{tabular}
\caption{A summary of community decisions listed across the four workshops. Participants noted that a community's specific goals  shaped all of these decisions.}
\Description{Groupings of community decisions mentioned in the workshops. The decisions are in three groupings: technical and infrastructural decisions; organizational decisions; and normative decisions. Technical infrastructural decisions contain things like: relay, server and other hosting infrastructure, platform software, tools like bots, member-facing options. Organizational decisions contain things like: number of admins, ownership, legal obligations, decision-making model and processes. Normative decisions have things like: federation policy, terms of service, bridging, blocklist, data permanence and maintenance.}
\label{tbl:decisions}
\end{table*}

\begin{table*}[h]
\centering
\begin{tabular}{|p{2cm}|p{3.65cm}|p{3cm}|p{3cm}|p{3.5cm}|}
\hline
\textbf{Concept} & \textbf{W1} & \textbf{W2} & \textbf{W3} & \textbf{W4} \\ \hline
Nutrition Facts &  & shared vocabulary of governance & encoded, standardized code of conduct; forking; mapping compatibility across communities & guide of community governance with a nutrition label \\ \hline
Trust Bubbles & RSS feed for a circle of communities & an ``archipelago'' of multiple coalitions; local bubbles and ``community of communities'' & sharing moderation tasks within a bubble, ideally integrated into existing software & ``ladder of engagement'' based on a ``chain of trust'' that de-dichotomizes the concept of federation \\ \hline
Shared Logs & moderation and blocking decisions in a public log archive, possibly represented as abstractions of patterns that need review & mechanism for pooling moderation tasks & moderator dashboard briefly mentioned &  \\  \hline
Issue Bulletin & meta-community; shared console, using algorithmic sorting to highlight trending and relevant discussions & a space for bouncing ideas and having discussions about issues & mod-mod communication tools, like a moderator inbox & sharing, verifying, collating, and triaging complaints and evidence of harm \\ \hline
\end{tabular}
\caption{High-level summary of how concepts from the ecosystem relate to more substantive discussions across the workshops (W\#), briefly noting the more specific or particular ways participants talked about the concept. Note that not all concepts were explored equally in depth in every workshop, and all of the concepts are somewhat tied to each other.}
\label{tbl:tool_ideas}
\end{table*}

\end{document}